\documentclass[fleqn,usenatbib]{mnras}

\usepackage[T1]{fontenc}
\usepackage{ae,aecompl}

\usepackage{graphicx}	
\usepackage{amsmath}	
\usepackage{amssymb}	

\usepackage{newtxtext,newtxmath}

\title[A variable corona for GRS 1915$+$105]{A variable corona for GRS 1915$+$105 }

\author[K. Karpouzas et al.]{
Konstantinos Karpouzas$^{1,2}$\thanks{E-mail: karpouzas@astro.rug.nl}, Mariano M\'endez$^{1}$, Federico Garc\'ia$^{1}$, Liang Zhang$^{2}$,
\newauthor
\ Diego Altamirano$^{2}$, Tomaso Belloni$^{3}$ and Yuexin Zhang$^{1}$
\\
$^{1}$Kapteyn Astronomical Institute, University of Groningen, P.O. BOX 800, 9700 AV Groningen, The Netherlands\\
$^{2}$School of Physics and Astronomy, University of Southampton, Southampton, SO17 1BJ, UK\\
$^{3}$INAF - Osservatorio Astronomico di Brera, via E. Bianchi 46, I-23807 Merate, Italy
\\
}

\date{Accepted 2021 March 16. Received 2021 March 08; in original form 2020 October 04}

\pubyear{2020}

\begin{document}
\label{firstpage}
\pagerange{\pageref{firstpage}--\pageref{lastpage}}
\maketitle

\begin{abstract}
Most models of the low frequency quasi periodic oscillations (QPOs) in low-mass X-ray binaries (LMXBs) explain the dynamical properties of those QPOs. On the other hand, in recent years reverberation models that assume a lamp-post geometry have been successfull in explaining the energy-dependent time lags of the broad-band noise component in stellar mass black-holes and active galactic nuclei. We have recently shown that Comptonisation can explain the spectral-timing properties of the kilo-hertz (kHz) QPOs observed in neutron star (NS) LMXBs. It is therefore worth exploring whether the same family of models would be as successful in explaining the low-frequency QPOs. In this work, we use a Comptonisation model to study the frequency dependence of the phase lags of the type-C QPO in the BH LMXB GRS 1915$+$105. The phase lags of the QPO in GRS 1915$+$105 make a transition from hard to soft at a QPO frequency of around 1.8 Hz. Our model shows that at high QPO frequencies a large corona of $\sim$ 100$-$150 $R_g$ covers most of the accretion disc and makes it $100\%$ feedback dominated, thus producing soft lags. As the observed QPO frequency decreases, the corona gradually shrinks down to around 3$-$17$R_g$, and at 1.8 Hz feedback onto the disc becomes inefficient leading to hard lags. We discuss how changes in the accretion geometry affect the timing properties of the type-C QPO.

\end{abstract}




\section{Introduction}
Low frequency (LF) Quasi-Periodic Oscillations (QPOs) in stellar-mass black hole binaries (BHB) have been known for many years (see reviews by \citealt{Motta2016} and \citealt{Ingram2020}). These QPOs are distinct peaks in the Power Density Spectra (PDS) of X-ray light curves of these sources. Based on the strength of the underlying broad-band variability, centroid frequency, $\nu$, and quality factor, $Q=\nu/\Delta \nu$, where $\Delta \nu$ is the full width at half maximum around the centroid frequency, LF QPOs are divided into three types, type-A, -B and -C (\citealt{Casella2004A&A}). While all of these types share similar centroid frequencies, type-C are the most frequent QPOs, with high $Q$-factors. Although type-C QPOs usually appear in the range of a few mHz to 10 Hz, they have also been detected at frequencies as high as 30 Hz (\citealt{Revnivtsev2000MNRAS}). Models that explain the dynamical origin of QPOs in BHBs were proposed about two decades ago when \cite{Stella1998} introduced the idea of relativistic Lense-Thirring precession (LTP) as the underlying mechanism. Follow-up theoretical work by \cite{Stella1999ApJ}, \cite{Psaltis2000}, \cite{Fragile2001}, \cite{Schnittman2006ApJ} and \cite{Ingram2009MNRAS} connected the LF QPOs in BHBs to the LTP frequency. Since then, there have been a plethora of studies that used timing and spectral features to add support to LTP as the origin of LF QPOs (\citealt{Motta_b_2014MNRAS}; \citealt{Motta_a_2014MNRAS}; \citealt{Ingram2016} and \citealt{Ingram2017}). Other models, implementing different physical mechanisms were being developed almost at the same time; some of those are the Two-Component Advection Flow model (TCAF, \citealt{Molteni1996ApJ}; \citealt{Chakrabarti2008A&A}) and the Accretion-ejection instability model (\citealt{Tagger1999}). For recent reviews of observations and theory of LF QPOs, we refer the reader to \cite{Done2007A&ARv}, \cite{Motta2016} and \cite{Ingram2020}.

The picture of the geometry of accretion onto a BH presented in \cite{Ingram2009MNRAS} consists of two main components. The first component is the classical optically thick geometrically thin accretion disc (\citealt{Shakura1973}), which is thought to be truncated at a radius $r_t$ from the BH, typically larger than the inner-most stable circular orbit (ISCO), $r_{ISCO}$, depending on the spectral state (\citealt{Ichimaru1977ApJ}; \citealt{Esin1997ApJ}; \citealt{Poutanen1997MNRAS}; \citealt{Gierli2008};). The second component is a geometrically thick hot flow that is assumed to reside between the truncated disc and the ISCO, and to be misaligned with respect to the accretion disc plane. General relativistic hydrodynamical simulations suggest that, under certain conditions, such thick inner flows can occur (\citealt{Fragile2005ApJ}; \citealt{Fragile2007ApJ}). The core idea of \cite{Ingram2009MNRAS} is that the type-C QPO is produced by such torus-like inner flow which precesses at the LTP frequency. The outer radius of the torus, $r_o$, is smaller or equal to the truncation radius of the accretion disc, while the inner radius is set by the limit at which the surface density of the torus becomes sufficiently low, due to the high tilt angle of its inner parts (see \citealt{Ingram2009MNRAS} for a detailed description). The idea of a precessing torus, rather than a precessing disc, is partly motivated by the fact that the energy spectrum that is modulated at the frequency of the type-C QPO (covariance or rms spectrum), is well described by a Comptonised spectrum (\citealt{Sobolewska2006MNRAS}).

\par The general spectral behaviour of BHBs is usually described by a transition from a hard, less luminous, to a soft luminous state, imprinted as a q-shaped trajectory on the Hardness-Intensity Diagram (HID; \citealt{Belloni2005A&A} and \citealt{Homan2005Ap&SS}), with regimes of intermediate hardness and luminosity. The frequency of the type-C QPO increases as the source moves from harder to softer states. Depending on the model, the change in hardness in the HID can be interpreted as a change of the outer radius of a hot inner flow (\citealt{Ingram2009MNRAS}) or, for models with an extended Comptonising medium (\citealt{Kazanas1997ApJ}), as a change in properties of this Comptonising medium such as size, optical depth and temperature. Hereafter, we will refer to this extended Comptonising medium as the corona (\citealt{Thorne1975ApJ}; \citealt{Sunyaev1979Natur}).

\par The timing properties of the type-C QPO, namely the fractional rms amplitude in a broad energy band and phase lag between a hard and a soft band, have also been shown to depend upon QPO frequency and source inclination. In particular, \cite{Motta2015MNRAS} suggested that the rms amplitude of the type-C QPO as a function of QPO frequency is systematically higher for high than for low-inclination sources. Similarly, \cite{vde2017MNRAS} provided evidence that above a certain QPO frequency in low-inclination sources the phase lags of the type-C QPOs are positive (hard lags), meaning that high-energy photons arrive at the observer after the low-energy photons, and these hard lags increase with QPO frequency, whereas for high-inclination sources the phase lags are negative (soft lags) and decrease with QPO frequency. The aforementioned results favour a geometric origin for the type-C QPOs, which was further supported by more quantitative results, such as the modulation at half the QPO frequency of the centroid energy of the iron emission line in H$1743-322$ (\citealt{Ingram2016}). Many of the observational findings mentioned above still remain unexplained.

\par Over the years, considerable work has been done to study the dependence of the phase lags between two broad energy bands not only upon QPO frequency, as we mentioned in the previous paragraph, but also upon the frequency of the broad-band noise. At low Fourier frequencies of the broad-band noise, the lags are hard (\citealt{Miyamoto1988Natur}; \citealt{Kotov2001MNRAS}), and thought to be produced by fluctuations of the mass accretion rate that propagate from the outer to the inner part of the accretion disc (\citealt{Lyubarskii1997MNRAS}; \citealt{Arevalo2006MNRAS}; \citealt{Ingram2013MNRAS}),  and are imprinted on the Comptonised and un-Comptonised emission that is used to calculate these lags. At higher Fourier frequencies of the broad-band noise the magnitude of the lags decreases, and the lags usually become soft, providing strong evidence for reverberation (\citealt{Uttley2011MNRAS}; \citealt{DeMarco2015ApJ}). Strong reverberation signatures were also recently reported in the broad-band noise component of MAXI J$1820+070$ by \cite{Kara2019Natur}, who studied the soft phase lags of the broad-band noise both in the energy and frequency domain. These authors suggest that a compact corona with a height of about 5 $R_g$, where $R_g=GM/c^2$ is the gravitational radius, and $G$, $M$ and $c$ are the gravitational constant, BH mass and speed of light, respectively, illuminates an accretion disc that is truncated at $\sim$ 2 $R_g$, thus producing the measured soft lags. As the corona contracts, the soft lags become shorter due to the shorter light travel-time. Further attention to reverberation was given, since an accurate reverberation mapping model was developed by \cite{Ingram2019MNRAS} that is able to put constraints on the mass of the BH, as was shown by \cite{Mastroserio2019MNRAS}.

Undoubtedly, reverberation is an essential physical mechanism for modeling spectra of BHBs at certain states.  However, the treatment of the corona as a point-like source at a fixed height above the accretion disc in reverberation models limits the prospects of a self-consistent explanation of the observed phenomenology. In fact, several previous studies (\citealt{Miller2010MNRAS}; \citealt{Legg2012ApJ}; \citealt{Mizumoto2018MNRAS}; \citealt{Mizumoto2019MNRAS} have presented the timing analyses of Active Galactic Nuclei (AGNs) where, in general, the phenomenology is similar and requires the presence of an extended corona that partially covers the accretion disc. Moreover, the soft lags of the LF QPOs  in stellar mass BHBs, on which we focus from now on, are usually much larger than the lags of the broad-band noise component (\citealt{Wijnands1999ApJ}), and therefore a compact corona or hot flow with a small scale height is unable to produce such large soft lags. Specifically for stellar mass BHBs, \cite{Mizumoto2016PASJ} showed that in the case of GRS 1915$+$105 the lack of a variation of the iron line at different timescales further supports the idea of an extended corona. In addition, \cite{DeMarco2016ApJ} showed that in order to explain the large soft lags of H 1743$-$322 during the hard states, when the type-C QPO is present, with reverberation, one has to assume an extended corona as an illuminating source, with a height of a few hundred $R_g$. We note also that the broad-band noise component in NS and BHC sources consists of a combination of QPO-like components (\citealt{Nowak2000MNRAS}; \citealt{Belloni2002ApJ}) and therefore it is natural to study the rms and lag spectra of these QPO-like components to understand the properties of the boradband noise.

\par Although the soft lags of LF QPOs have been extensively studied in the literature, little work has been done to explain quantitatively their energy or frequency dependence in the context of a radiative mechanism. \cite{Nobili2000ApJ} suggested a Comptonisation model that consists of a two-component corona, which produces both hard and soft lags through Compton up- and down-scattering, respectively, in these two different components. Their model successfully explains both the magnitude and the change of sign of the lags as a function of QPO frequency in GRS 1915$+$105. 

\par Recently, \cite{Karpouzas2020} presented a model that explains the observed soft lags of the lower kHz QPO in neutron star (NS) LMXB, as a delayed heating of the seed photon source by photons previously up-scattered in the corona. This effect is referred to as feedback. In the case of NS LMXBs, it has been shown through Monte Carlo simulations (\citealt{Kumar2016MC}) that feedback should play an important role even among different corona geometries, and that feedback depends on the size of the corona. Although the geometry of the corona and seed photon source are very different between NS and BH LMXBs, one can in principle test how well feedback of up-scattered photons in an extended corona can explain the soft lags observed in BH LMXBs. In fact, based on the model of \cite{Karpouzas2020}, \cite{Garcia2021MNRAS} showed that a two-component extended corona explains the energy dependence of the time lags and rms amplitude of the type-B QPO in MAXI J1348$-$630, and so in this work we extend our analysis for type-C QPOs. A good candidate to perform such a test is GRS 1915$+$105 (\citealt{Castro1992}; \citealt{Castro1994}). GRS 1915$+$105 is one of the most observed, and thus best studied, sources in the Rossi X-ray timing explorer (\citealt{Bradt1993}, RXTE) archive. The compact object in GRS 1915$+$105, is a BH with a mass $M_{BH}=12.4^{+2}_{-1.8}$ $M_{\odot}$, (\citealt{Reid2014ApJ}) and spin estimated, with spectral techniques, to be between $a^*=0.68$ and $a^*=0.99$ (\citealt{spin2015}). 

\par GRS $1915+105$ is a peculiar source, in the sense that it never showed a full Q-shaped cycle on the HID. For a dedicated review on GRS 1915$+$105, we refer the reader to \cite{Fender2004MNRAS}. Its spectral behaviour was separated in 14 classes based on the work of \cite{Belloni2000} (see also \citealt{KW2002MNRAS} and \citealt{Huppenkothen2017MNRAS}). Timing studies of GRS 1915$+$105, which are mainly what we are interested in, date back to \cite{Xingming1997ApJ} and \cite{Morgan1997ApJ} and since then, several others focused on the spectral-timing properties of LF QPOs in this source (\citealt{Markwardt1999}; \citealt{Vignarca2003A&A}; \citealt{Rodriguez2004ApJ}). One of the most notable timing properties related to the type-C QPO phase lag between two broad bands of GRS 1915$+$105 is possibly the switch from hard lags, at QPO frequencies below $\sim$2 Hz (\citealt{Reig2000ApJ};\citealt{Pahari2013ApJ}; \citealt{Zhang2020MNRAS}), to zero at around 2 Hz, and then to soft lags at QPO frequencies above that limit. The slope of the lag-energy spectrum follows the same behaviour, changing from positive below 1.8 Hz to zero at around 1.8 Hz, and then negative above 1.8 Hz. Alongside with the phase lags, the rms amplitude, both energy- and frequency-dependent, of the type-C QPO also shows a particular dependence upon QPO frequency (\citealt{Zhang2020MNRAS}). 

Recent work from \cite{Dutta2016ApJ} and \cite{Chatterjee2017MNRAS}, implementing the TCAF model (\citealt{Chakrabarti2008A&A}), linked the QPO frequency to the size of the Centrifugal pressure supported Boundary Layer (CENBOL), which serves as the corona in the TCAF model. These studies suggested that changes in the size of the CENBOL can explain the change in sign of the time lag. More recently, \cite{Dutta2018MNRAS} applied the TCAF model to data of GRS 1915$+$105, and showed how a gradual change in the size of the CENBOL can explain the time evolution of the type-C QPO frequency, above and below the 1.8 Hz limit. \cite{Dutta2018MNRAS} concluded that, as the QPO frequency decreases from over 5 Hz to below 1.8 Hz, the CENBOL increases in size from around 135 $R_g$ to over 500 $R_g$, and then it decreases again to about 100 $R_g$, when the QPO frequency increases beyond 5 Hz. 

\par Here, we use the Comptonisation model of \cite{Karpouzas2020} to explain the dependence of both the rms amplitude and phase lags upon QPO frequency and energy. In sections 2 and 3 we, respectively, introduce the model and explain its application to the data of \cite{Zhang2020MNRAS}. In section 4 we show our results. Finally, in section 5 we discuss our results and provide a new view to how the corona evolves over time, and how the latter affects the timing behaviour of GRS 1915$+$105.

\section{Comptonisation model}

In this work we use the model presented by \cite{Karpouzas2020}, which is an adaptation of the Comptonisation model proposed by \cite{Lee1998}; \cite{Lee2001}; and \cite{Kumar2014}. The main idea behind these models is that any QPO can be described as an oscillation of the time averaged spectrum, $n_{\gamma0}$, that is coupled to oscillations of other properties of the system, such as the corona temperature, $kT_e$, seed-photon source temperature, $kT_s$, and the external heating provided to a finite-sized corona, $\dot{H}_{ext}$. No assumption is made about the origin of the QPO frequency in the model. As far as the model is concerned, the QPO is an oscillation of the X-ray flux available for Compton up-scattering in the corona. As a consequence, the properties of the corona and seed-photon source can react to the QPO and oscillate coherently. Another possibility is that the QPO itself is produced by some instability in $T_e$ and $\dot{H}_{ext}$ or some oscillatory mode in the corona (see \citealt{Ingram2009MNRAS}; \citealt{Fragile2016MNRAS}; \citealt{Fragile2020MNRAS}). 

\par The model of \cite{Karpouzas2020}, assumes that  the seed photon source is a blackbody and that the Comptonising corona is a spherically symmetric homogeneous shell with optical depth $\tau$ and thickness $L$ around that blackbody. The model also takes into account feedback of up-scattered photons onto the blackbody. When applying the model in practice, first we calculate $n_{\gamma0}$ by solving the Kompaneets equation (\citealt{kompaneets1957}) in steady-state. Next, we linearise the Kompaneets equation assuming that $kT_e$, $kT_s$ and $\dot{H}_{ext}$ undergo small oscillations at exactly the QPO frequency, and solve for the complex amplitude, $\delta n_{\gamma}$, of the averaged spectrum, $n_{\gamma 0}$. We note that the oscillation of the seed-photon source temperature, $kT_s$, is attributed to feedback photons. Feedback, within the model, is regulated by the feedback fraction, $f_{\eta}$, that is defined as the fraction of the luminosity of the seed source that is solely due to feedback. The complex amplitude, $\delta n_{\gamma}$, holds information about the energy-dependent amplitude and phase lags at the QPO frequency, $\nu_{QPO}$, that was used to perform the linearisation.  We note here that only $n_{\gamma0}$ and $\delta n_{\gamma}$ are functions of photon energy, while $kT_e$, $kT_s$, $L$, $\tau$ and $f_{\eta}$ are considered constant for a specific QPO frequency. In \cite{Karpouzas2020} we showed that the model can fit the energy-dependent time lags and fractional rms amplitude of the kHz QPOs of the NS LMXB, 4U $1636-53$. In this paper we make a first attempt to quantitatively explain the timing properties of the type-C QPO in the BH LMXB GRS 1915$+$105 using this same model. Our approach introduces two caveats. Firstly, the source of seed photons that we use is a blackbody instead of a disc-blackbody. Secondly, the photon-loss term in the Kompaneets equation assumes an equal optical depth seen by all seed-photons, which is a crude approximation in the case of a disc whose spatial extent is comparable to the size of the corona. We will expand on these issues in the Discussion section.

\section{Data Analysis and Modeling}

\cite{Zhang2020MNRAS} presented an extensive timing analysis of 620 observations of GRS 1915$+$105 showing type-C QPOs, after analysing all of the RXTE archival data (1996$-$2012). The QPO frequencies ranged from 0.4 Hz to 6.3 Hz. \cite{Zhang2020MNRAS} measured the phase lags of all type-C QPOs between two broad bands, 2$-$5.7 keV and 5.7$-$15 keV. They also measured the lags between multiple narrow energy bands, which we hereafter refer to as lag-energy spectra. In the paper of \cite{Zhang2020MNRAS}, lag-energy spectra at nine selected QPO frequencies were shown for reference. Similarly, \cite{Zhang2020MNRAS} measured the fractional rms amplitude both in the full PCA band (2$-$60 keV), and in separate bands. The latter will be referred to as the rms-energy spectrum. The primary focus of  \cite{Zhang2020MNRAS} was to study the frequency dependence of the phase lags between the two broad bands mentioned above, which they called the lag-frequency spectrum. The authors fitted a broken line to the lag-frequency spectrum and placed a sufficiently precise limit of $1.8 \pm 0.1$ Hz on the QPO frequency at which the measured phase lags change from hard to soft. Their best-fitting broken-line to the lag-frequency spectrum also exhibited a statistically significant difference in slope, with a value of  $-0.21 \pm 0.02$ below 1.8 Hz and $-0.1 \pm 0.01$ above 1.8 Hz, indicating a potential change in the physical mechanism that produces the lags.

\par \cite{Zhang2020MNRAS} showed that the frequency dependence of the rms amplitude in the full PCA band, which we will refer to as the rms-frequency spectrum, also changes behaviour around the same QPO frequency, switching from increasing with QPO frequency for QPO frequencies below 1.8 Hz, to decreasing for QPO frequencies above 1.8 Hz. Although the model of \cite{Karpouzas2020} can predict both energy and frequency dependence of the fractional rms amplitude and phase lags, in this work we first study the frequency-dependent fractional rms amplitude in the full band, and phase lags between the two broad bands used in \cite{Zhang2020MNRAS}. The latter approach allows us to analyze all 620 QPO frequencies, while avoiding the computationally expensive process of fitting the rms- and lag-energy spectra at all QPO frequencies. By studying the frequency dependence, we obtain an initial estimate of how the model parameters depend upon QPO frequency. Then we proceed with fitting the rms- and lag-energy spectra presented by \cite{Zhang2020MNRAS}, at representative QPO frequencies, and compare with our initial estimate for completeness. The challenge of the frequency-dependent analysis, with which we start, is that we only have two measurements at our disposal, namely the fractional rms amplitude and phase lag for each set of the 7 parameters of the model, $kT_e$, $kT_s$, $\tau$, $L$, $f_{\eta}$, $\delta \dot{H}_{ext}$ that are needed to produce a pair of simulated fractional rms amplitude and phase lag at a certain QPO frequency, $\nu_{QPO}$. Thus, one can not perform a fitting in the classical sense. In the following, we describe how we can overcome the aforementioned limitation by matching the predicted, from the model, frequency dependence of the fractional rms amplitude and phase lags to the measured ones. Hereafter, we will refer to the term fractional rms amplitude simply as rms amplitude.

\par As shown by \cite{Karpouzas2020}, for a given set of the 7 parameters mentioned above, the model simultaneously predicts the rms amplitude in any band and phase lag between any two bands. However, the amplitude of the external heating rate, $\delta \dot{H}_{ext}$, only affects the rms amplitude, acting as a normalisation, while the phase lags are not dependent on this parameter. Therefore, a good approach is to only simulate the lag-frequency spectrum at first, and then re-scale the simulated rms-frequncy spectrum, that was anyway produced alongside the lag-frequency spectrum, by adjusting $\delta \dot{H}_{ext}$ in order to match the measured rms amplitude values. In the following sections we discuss the modeling of the phase lags first, and then we study the resulting rms amplitude.

\subsection{Modeling of the lag-frequency spectrum}
\label{sub:lag-frequency}

\par To explain the change of sign of the phase lag around 1.8 Hz and its dependence upon QPO frequency, we designed a simple computational experiment using the model of \cite{Karpouzas2020}. In particular, we generated $10^6$ random combinations of the model parameters uniformly for $kT_e$ in the range 3$-$100 keV, $kT_s$ in the range 0.1$-$3 keV, $L$ in the range 2$-$500 $R_g$, $f_{\eta}$ in the range 0$-$1 and $\nu_{QPO}$ in the range 0.4$-$6.3 Hz. We then calculated $\tau$ using $kT_e$ and a random uniformly distributed value of $\Gamma$ between 1.5 and 3, to cover a wide enough range of the power-law index. To calculate $\tau$, for every selected value of $kT_e$ and $\Gamma$, we used the formula:

\begin{equation} \label{eq:tau}
\tau  = \sqrt{2.25 + \frac{3}{ \frac{kT_e}{m_ec^2}\big[ (\Gamma + 0.5)^2 - 2.25 \big]}},
\end{equation}

where $m_e$ is the electron mass. We set the external heating rate, $\delta \dot{H}_{ext}$, to a constant arbitrary value of $1\%$, since it only affects the measured rms amplitude and, as has been shown by \cite{Karpouzas2020}, it has a dependence upon QPO frequency, which we do not know a priori. Later, when we compare with the measured rms amplitude, we can re-scale $\delta \dot{H}_{ext}$ to match the data, and reveal its QPO frequency dependence and actual value. 

\par
\begin{figure}
\hspace*{-1.5cm}
	\includegraphics[width=\columnwidth,angle =-90]{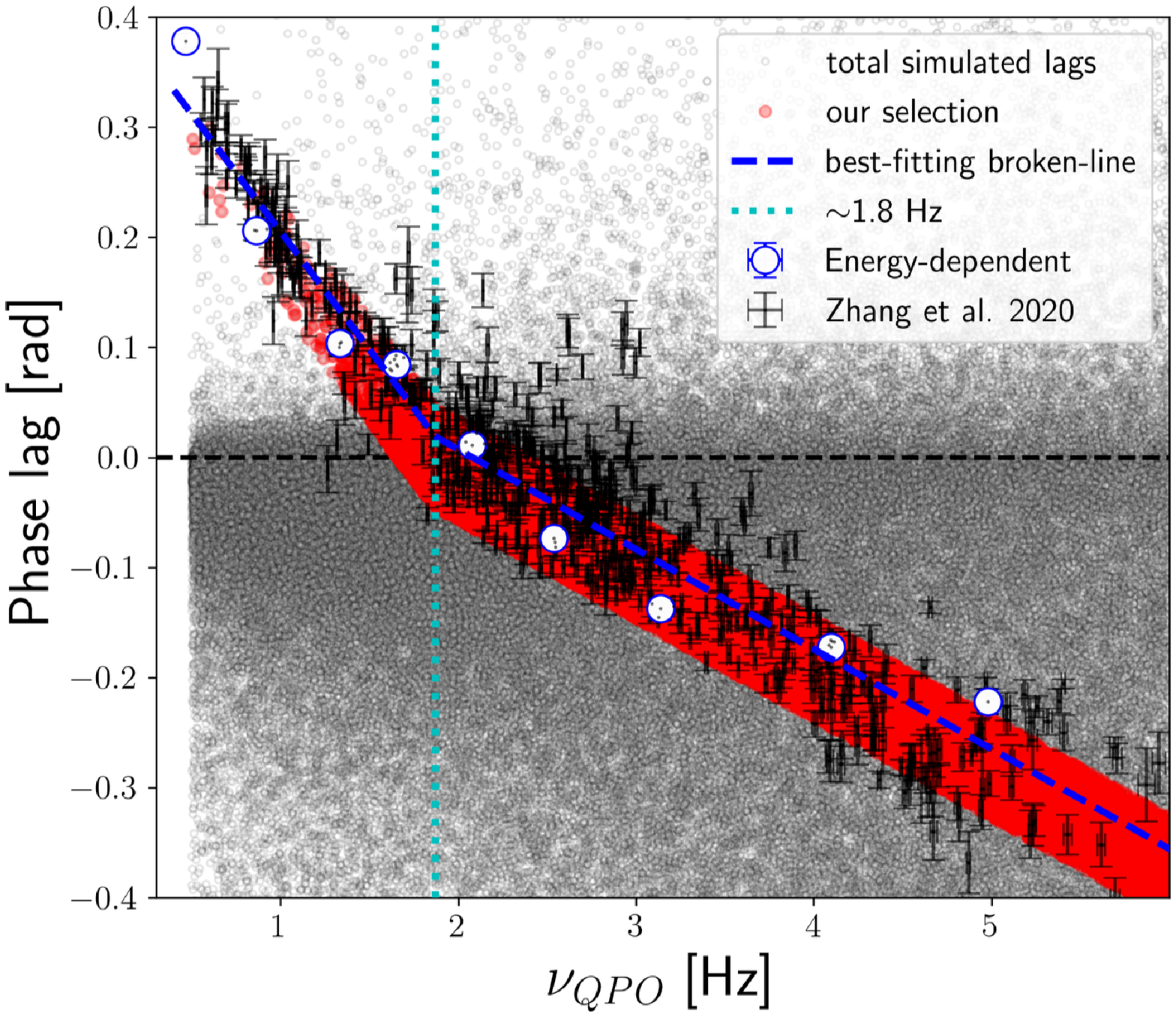}
    \caption{The measured lag-frequency spectrum of GRS 1915$+$105 (black points with error bars) alongside with all of our simulations. Grey circles represent all of the simulated pairs of phase lags and QPO frequencies. Red represents the simulated pairs that cluster within 1-$\sigma$ of the best-fitting broken line (blue dashed line) to the measured phase lags. The break, where the phase lags switch from hard to soft, is denoted by the cyan dotted line at $\sim$1.8 Hz. The white circles that track the best-fitting broken line show individual observations, at selected QPO frequencies, for which the energy dependence of the phase lags was measured (see section \ref{subsection:MCMC}).} 
    \label{fig:plag_clustering}
\end{figure}

 For each randomly generated QPO frequency, there is a corresponding randomly generated combination  of the five model parameters $kT_s$, $kT_e$, $L$, $f_{\eta}$ and $\tau$, that when given as input to the model generates a model estimate of the phase lag between the bands 2$-$5.7 keV and 5.7$-$15 keV, at that QPO frequency. The energy bands were chosen such that we maintain consistency with \cite{Zhang2020MNRAS}. If one plots all the values of the model estimated phase lags as a function of the randomly generated QPO frequency, as we have done with gray circles in Figure \ref{fig:plag_clustering}, no obvious correlation between the two is observed. However, the data suggests that there exists a significant correlation between phase lag and QPO frequency. To accommodate for the observed correlation, we created two sub-samples from our total simulated sample of phase lags, one with pairs of simulated phase lags and QPO frequencies for which the phase lags where positive, at QPO frequencies below 1.8 Hz and clustered around the best-fitting broken line of \cite{Zhang2020MNRAS} below 1.8 Hz, and another with pairs of simulated phase lags and QPO frequencies for which the phase lags where negative, at QPO frequencies above 1.8 Hz and clustered around the best-fitting broken line above 1.8 Hz. As a condition for the clustering around the best-fitting broken line, we demanded that the predicted phase lag from the model can deviate from the best-fitting broken line as much as one standard deviation of the data. The aforementioned computational experiment is illustrated in Figure \ref{fig:plag_clustering}, alongside with the original data and best-fitting broken line of  \cite{Zhang2020MNRAS}.

\par For each pair of simulated phase lag and QPO frequency, that we restricted alongside the best-fitting broken line in Figure \ref{fig:plag_clustering}, there is also a pair of the five remaining model parameters $kT_s$, $kT_e$, $L$, $f_{\eta}$ and $\tau$. Our goal was to check whether the aforementioned model parameters, that were used to generate the clustered simulated points around the broken line, have any correlation with QPO frequency that separates them from the total simulated population which is uniformly sampled. We will discuss the dependence of the model parameters upon QPO frequency in the Results section of this paper, where they are presented, but for now we will discuss the other by-product of our model, the rms amplitude, which at this point is already calculated but not compared with the measured rms amplitude.

\subsection{The rms-frequency spectrum}

\begin{figure}
\hspace*{-1.5cm}
	\includegraphics[width=\columnwidth,angle =90]{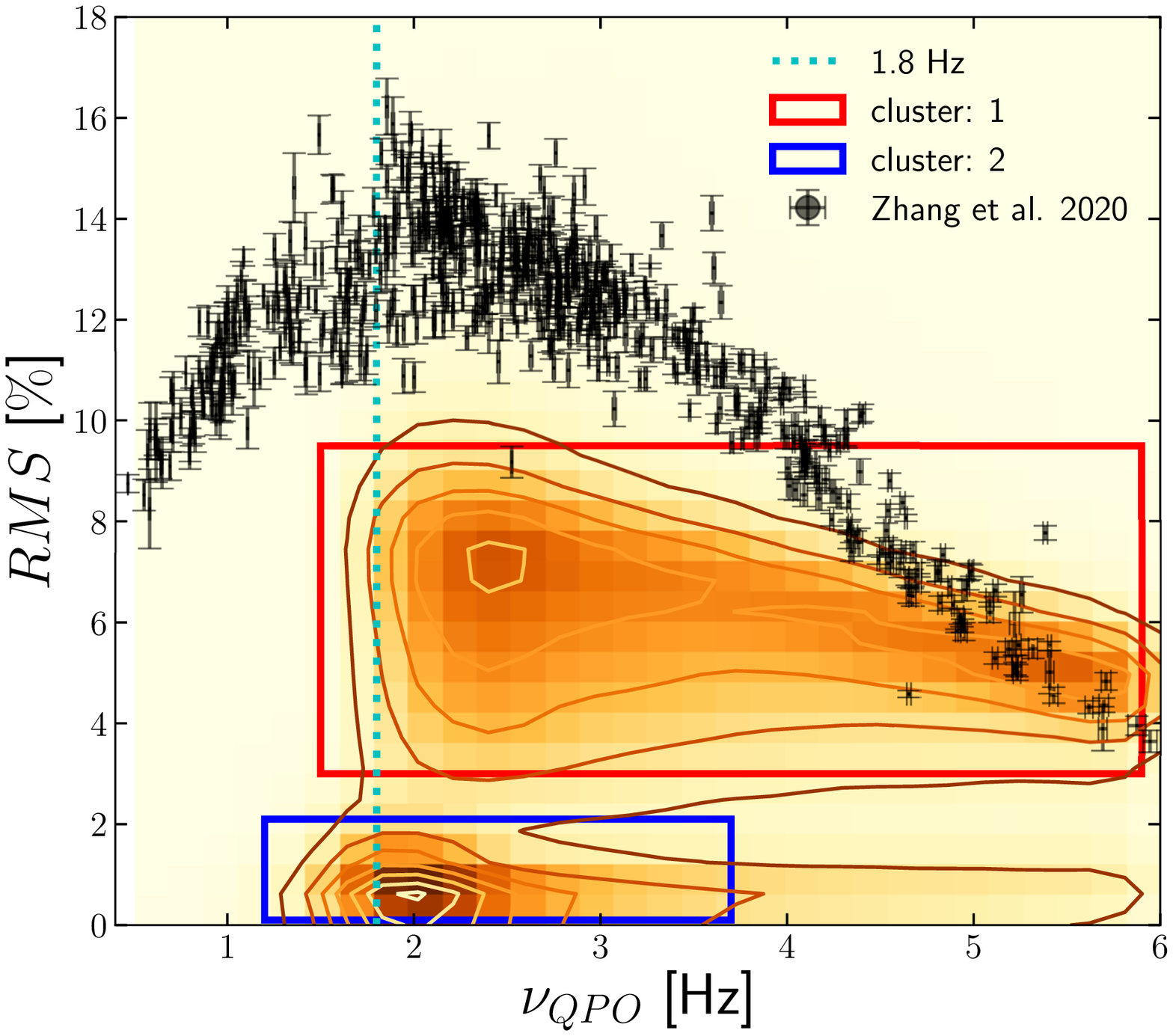}
    \caption{The measured rms-frequency spectrum of GRS 1915$+$105 (black points with errorbars) alongside with the simulated rms-frequency pairs that follow the broken line of Figure \ref{fig:plag_clustering} (contours). The contours were produced by smoothing the two-dimensional space of the simulated points with a Gaussian kernel. The rms amplitude is re-scaled by a factor of $11 \%$ to better compare with the data. The red and blue rectangles showcase the separation of the two prominent clusters in the simulations.  }
    \label{fig:rms_clustering}
\end{figure}

As we mentioned above, our model can predict the rms amplitude at any QPO frequency, simultaneously with the phase lag, for any combination of the model parameters. However, we remind the reader that to compare with measured rms amplitude we need to take into account the amplitude of the external heating rate , $\delta \dot{H}_{ext}$. Up to this point, we have clustered the simulated pairs of phase lags and frequencies alongside the best-fitting broken line (Figure \ref{fig:plag_clustering}). In Figure \ref{fig:rms_clustering}, we plot the simulated rms-frequency spectrum, i.e. the pairs of simulated rms amplitude and frequency, only for the points for which the corresponding simulated phase lags are clustered around the broken-line of Figure \ref{fig:plag_clustering}. On top of that, we plot the rms amplitude measurements of \cite{Zhang2020MNRAS} alongside with their uncertainties. In Figure \ref{fig:rms_clustering}, we re-scaled the simulated rms-frequency spectrum by multiplying all of the values by a factor of 11, which is equivalent to assuming an external heating rate amplitude of $\delta \dot{H}_{ext}=11\%$ instead of $1\%$ that we used before, equal at all QPO frequencies. This re-scaling is totally arbitrary and only serves as a way to visually compare the simulated rms-frequency spectrum to the measured values of \cite{Zhang2020MNRAS}.

\par One can immediately distinguish two clusters of simulated points in Figure \ref{fig:rms_clustering}. These clusters reflect the degeneracy caused by the lack of prior knowledge of the spectral parameters of the source in these observations (\citealt{Karpouzas2020}), $kT_e$, $\tau$ and $kT_s$. To detect and separate the clusters, we used the DBSCAN algorithm (\citealt{Ester1996ADA}). One of the clusters (blue rectangle in Figure \ref{fig:rms_clustering}) corresponds to very low values of the rms amplitude (< 1$\%$) and does not populate the higher QPO frequencies (>3 Hz) very well, even when arbitrarily re-scaled. The other cluster (red rectangle in Figure \ref{fig:rms_clustering}) appears at higher rms amplitude but does not fully populate the lower QPO frequencies (<2 Hz) sufficiently well. Interestingly, by forcing the simulated lag-frequency spectrum to follow the broken line in Figure \ref{fig:plag_clustering}, the corresponding simulated rms-frequency spectrum for the upper cluster (red rectangle in Figure \ref{fig:rms_clustering}), has a significant negative correlation, exactly above the 1.8 Hz level, as observed in the measured rms amplitude.

\par
\begin{figure}
\hspace*{-1.7cm}
	\includegraphics[width=\columnwidth,angle =90]{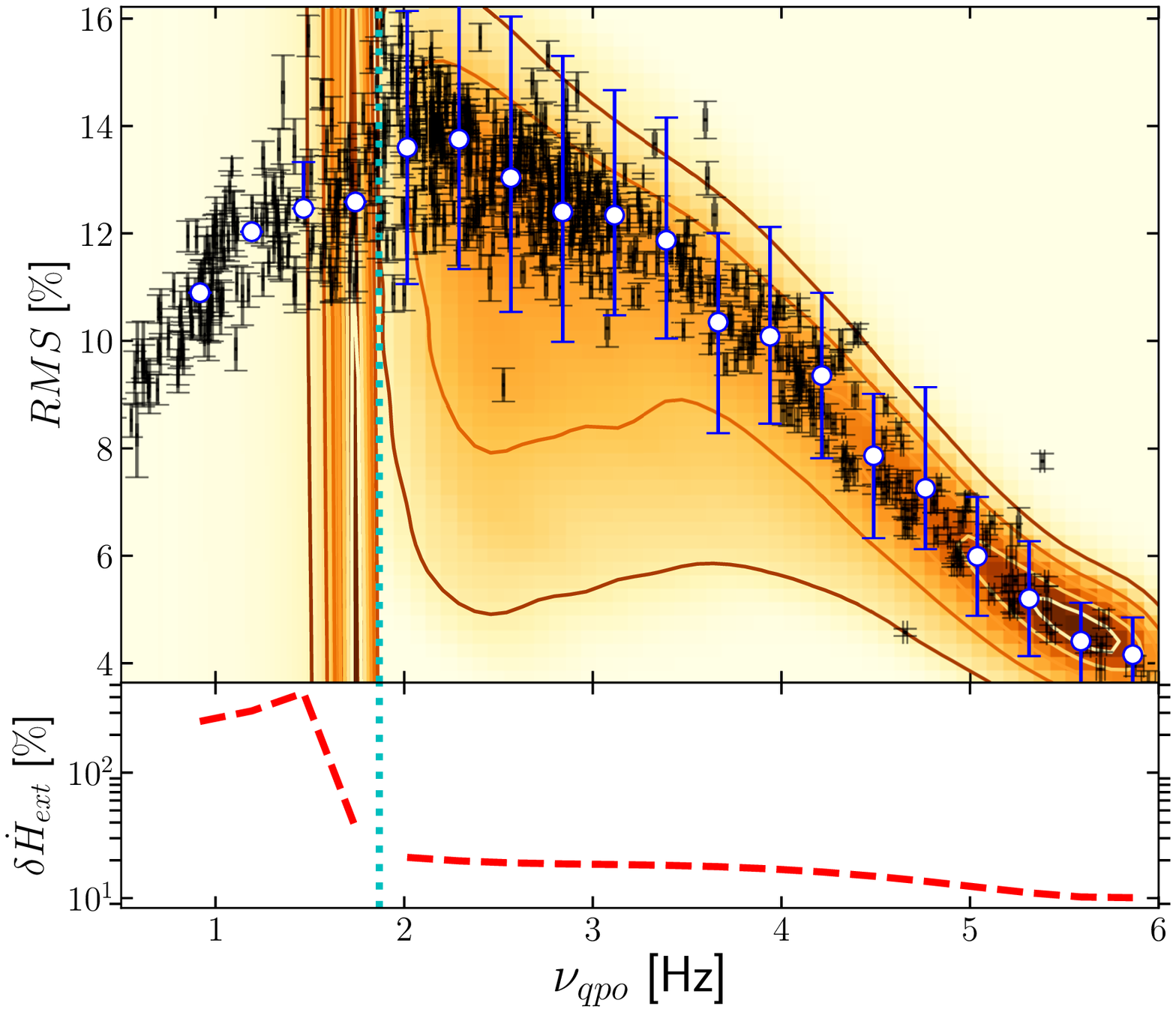}
    \caption{Matching of the simulated rms-frequency spectrum to the measured one for the type-C QPOs in GRS 1915$+$105. The top panel shows the data (black error-bars) alongside with the full simulated values of the rms amplitude (contours) which have been averaged in selected frequency bins (white circles). The blue error-bars indicate the 1-$\sigma$ interval around the mode of each distribution of simulated rms amplitude in a particular frequency bin. The red dashed line in the lower panel is the required external heating rate amplitude, at each QPO frequency bin, that minimises the residuals between the white circles (top panel) and the average measured rms (back errors).   }
    \label{fig:rms_sim_DHext}
\end{figure}

 Since $\delta \dot{H}_{ext}$ is QPO frequency-dependent, in general, the simulated rms-frequency spectrum can not perfectly match the data yet, for any arbitrary re-scaling factor that is constant at each frequency. However, if one matches the measured rms-frequency spectrum with the simulated one, the expected frequency dependence of $\delta \dot{H}_{ext}$ can be extracted. To perform the aforementioned matching, first we divided the simulated frequency range in 15 bins. In each bin we calculated the mode of the distribution of the simulated rms amplitude values and used that as the expected value of the simulated rms amplitude in that bin. We also calculated the $1-\sigma$ confidence interval around the mode, and used that value as the uncertainty of the simulated rms amplitude in that bin. After doing that, we have an rms-frequency relation which we can match to the measured rms-frequency relation. To do the latter, we simply found a normalisation, different at each QPO frequency that, when multiplied with the simulated rms amplitude, re-produces the measured rms amplitude at each QPO frequency. The value of the normalisation factor at each QPO frequency represents the required external heating rate fractional amplitude, $|\delta \dot{H}_{ext}|$, that best fits the data. We plot the result of this matching in the top panel of Figure \ref{fig:rms_sim_DHext}, while the lower panel shows the values of the normalisation factor required at every QPO frequency.

\par We must note that we discarded the lower cluster (blue rectangle in Figure \ref{fig:rms_clustering}) of the simulated rms-frequency spectrum because it would require an extreme variability of the external heating rate ($\delta \dot{H}_{ext} \sim 100\%$), to agree with the measured rms amplitude. In Figure \ref{fig:rms_clustering}, the contour levels correspond to a smoothing of the simulated rms-frequency pairs in the two-dimensional space using a Gaussian kernel. Seemingly, the contours in Figure \ref{fig:rms_clustering} show no simulated rms-frequency pairs below 1 Hz. This happens only because the model, under the priors assumed, predicts too few points below 1 Hz, compared to the ones above that frequency, that they appear insignificant after the smoothing process, and thus not visible. This effect will be discussed below. 

\par To study the physical parameters of the model and their dependence upon QPO frequency, which is discussed in the next section, we separated the simulated QPO frequency domain in two regimes, one below 1.8 Hz and the other above 1.8 Hz. For the regime above 1.8 Hz, we used all the simulated models that belonged to the upper cluster of the rms-frequency spectrum (red rectangle of Figure \ref{fig:rms_clustering}). For the regime below 1.8 Hz, we used all the available simulated models due to the fact that the simulations in this regime were more scarce. The issue of scarce model sampling below 1.8 Hz will be addressed in Section \ref{s:caveats}.

\subsection{MCMC fitting to the rms- and lag-energy spectra}
\label{subsection:MCMC}

\cite{Zhang2020MNRAS} presented rms- and lag-energy spectra of GRS 1915$+$105 for selected type-C QPO frequencies in the complete range 0.4$-$6.3 Hz. Those selected frequencies are plotted as white circles in Figure \ref{fig:plag_clustering}, and can be viewed as snapshots of the timing properties of GRS 1915$+$105 at a given QPO frequency. They showed that, on average, the initially negative slope of the lag-energy spectra systematically increases following a linear trend, while the QPO frequency decreases down to $\sim$ 2 Hz. Below 2 Hz, the slope of the lag-energy spectrum, on average, crosses the zero level (switching to hard lags) and continues to increase following a significantly different linear trend than the one formed above the 2 Hz limit. Up to this point, our analysis has omitted any information about the energy dependence of the rms amplitude and phase lags provided by the model. Nevertheless, we arrived at an initial estimate of how the physical parameters depend upon QPO frequency. The dependence of the physical parameters upon QPO frequency are shown as black circles in Figure \ref{fig:feta}.

\par
\begin{figure}
\hspace*{-1.99cm}
	\includegraphics[width=9cm,height=12cm,angle=90]{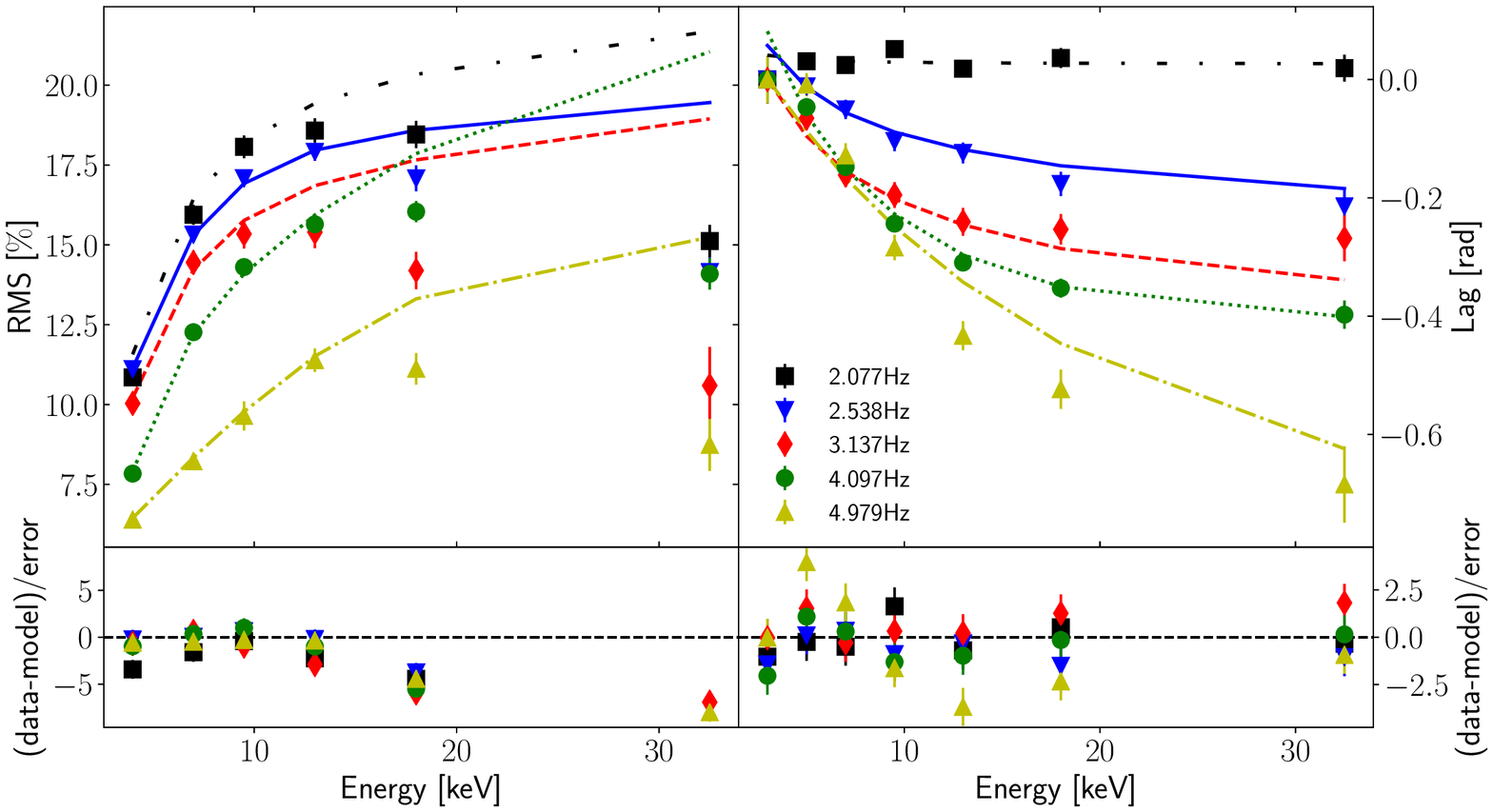}
    \caption{ Best-fitting energy-dependent models to the rms- and lag-energy spectra at selected type-C QPO frequencies above 1.8 Hz in GRS 1915$+$105. In the left and right panels the rms- and lag-energy spectra are plotted, respectively, alongside with the best fitting models. For the different QPO frequencies, the data-points and model-lines are plotted with different styles colors. In the bottom sub-panels of each side we plot the residuals between the data and model normalised by the uncertainties in the data.  }
    \label{fig:above}
\end{figure}

\begin{figure}
\hspace*{-1.99cm}
	\includegraphics[width=9cm,height=12cm,angle=90]{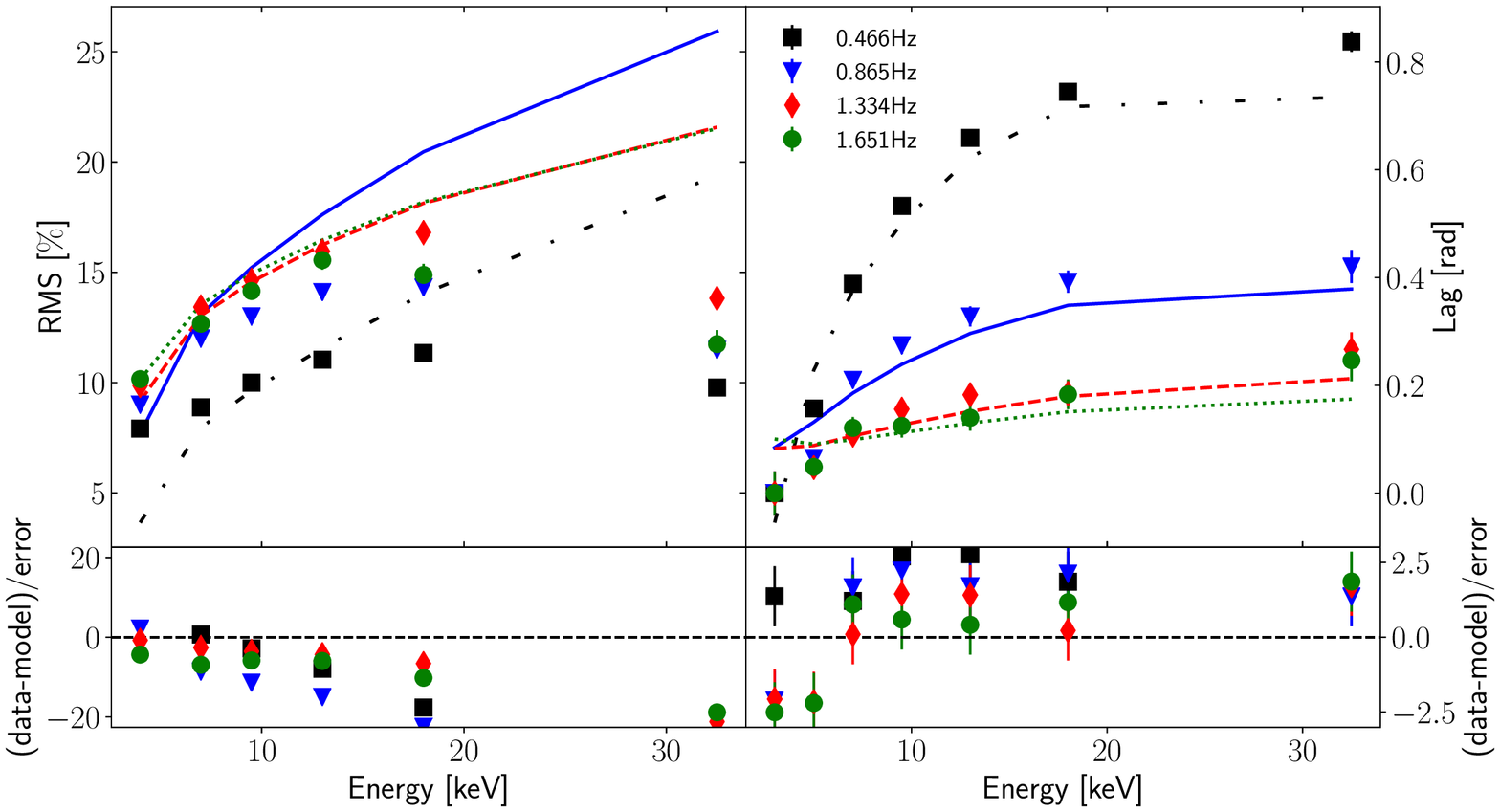}
    \caption{ Same as in Figure \ref{fig:above}, for the type-C QPO frequencies in GRS 1915$+$105 at QPO frequencies below 1.8 Hz. }
    \label{fig:below}
\end{figure}

 Using this initial estimate as a starting point, we performed an individual Markov-chain Monte Carlo (MCMC) fitting to each rms- and lag-energy spectrum selected by \cite{Zhang2020MNRAS}. For the MCMC fitting and analysis we used the affine-invariant ensemble sampler (\citealt{emcee2013PASP}) and followed the same approach as in \cite{Karpouzas2020}. We must note that in the case of the rms-energy spectrum, for reasons that we discuss in Section \ref{s:caveats}, we neglected the rms amplitude measurements above 25 keV. The best-fitting model, alongside with the data of \cite{Zhang2020MNRAS}, at each QPO frequency is plotted separately for the frequencies above 1.8 Hz in Figure \ref{fig:above} and for the frequencies below 1.8 Hz in Figure \ref{fig:below}. As expected, the MCMC fitting gives a separate estimate of the dependence of the physical parameters upon QPO frequency. In the next sections we will discuss these results and how they compare to the analysis that does not take the energy dependence of the rms amplitude and phase lags into account.

\section{Results}

\begin{figure*}
\hspace*{-3cm}
\centering
	\includegraphics[width=15cm,height=23cm,angle=90]{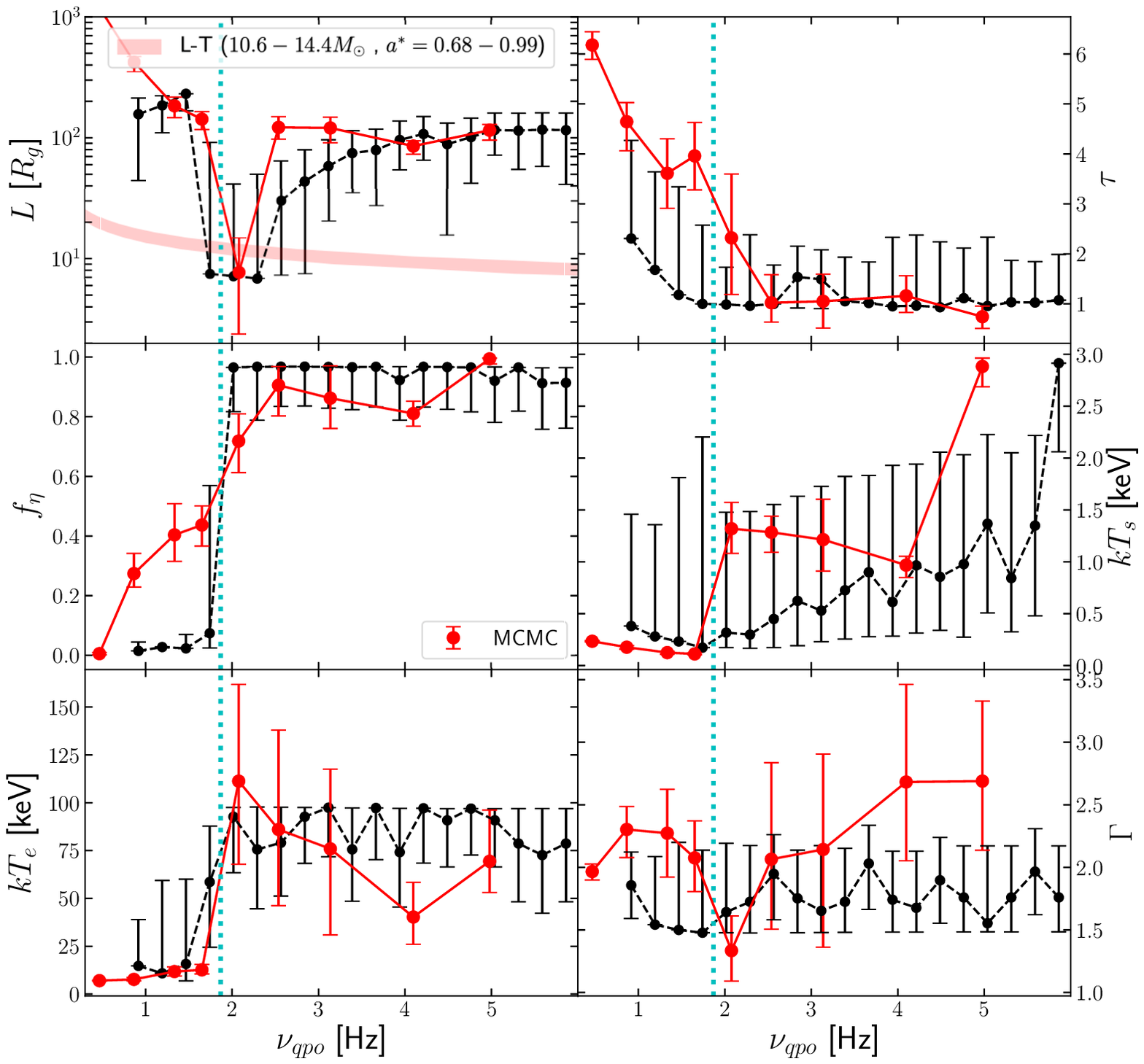}
    \caption{Dependence of the physical parameters upon the type-C QPO frequency in GRS 1915$+$105. In each panel the black circles show the parameter estimates based on the total QPO frequency sample (frequency-only-approach) while the red circles show the parameter estimates from the MCMC fitting applied to individual selected frequencies (energy-dependent-approach). The red shaded area, in the upper left panel shows the range of the predicted evolution of the disc inner radius assuming that the type-C QPO frequency is produced by LTP and assuming a range for the masses and spins of the BH in GRS 1915$+$105. The dotted cyan vertical lines, in all panels, denote the 1.8 Hz limit. }
    \label{fig:feta}
\end{figure*}

In section \ref{sub:lag-frequency}, we discussed the clustering of the simulated phase lags alongside the best-fitting broken-line of \cite{Zhang2020MNRAS}. This clustering provides an initial estimate of the frequency dependence of the model parameters, thus revealing what, in the context of the model, is the physical mechanism that drives the lag-frequency relation. In Figure \ref{fig:feta} we plot five of the model parameters, namely the corona size, $L$, feedback fraction, $f_{\eta}$, electron temperature of the corona, $kT_e$, optical depth, $\tau$, and seed photon source temperature, $kT_s$, as a function of the QPO frequency plus the photon power-law index, $\Gamma$, which is not an independent parameter but was used to generate the values of $\tau$ for the simulation (see section \ref{sub:lag-frequency}). 

\par To estimate the frequency dependence of the parameters, we divided the simulated frequency domain of the cases that clustered alongside the broken line in 15 bins. In each frequency bin we studied the distribution of each one of the five parameters and used the mode of the distribution as a best estimate, and the $1-\sigma$ confidence interval around that mode as the uncertainty of that estimate. These parameters, and their corresponding uncertainties at each QPO frequency are summarized in Table \ref{tab:frequency-only-approach}.
In Figure \ref{fig:feta} the black circles represent the values of the parameters as a function of QPO frequency derived from the clustering alongside the best-fitting broken-line (Figure \ref{fig:plag_clustering}). Hereafter, for simplicity, we will refer to the results derived from the clustering around the best-fitting broken-line process as the frequency-only-approach. The red circles in the same Figure represent the best-fitting parameters derived from the MCMC fitting to the individual rms- and lag-energy spectra of \cite{Zhang2020MNRAS} at selected QPO frequencies (see \citealt{Karpouzas2020} for details on the parameter estimation using the MCMC method). We will refer to the process of MCMC fitting to the rms- and lag-energy spectra as the energy-dependent-approach. The physical parameters derived from the energy-dependent-approach are summarized separately in Table \ref{tab:energy-dependent-approach}.

\par Based on the frequency-only-approach we find that the corona size decreases systematically from 115 $R_g$ at around 6 Hz down to about 7.2 $R_g$ at 2 Hz, and then increases again to almost 160 $R_g$ at 0.9 Hz. Alongside with the decreasing corona size, in Figure \ref{fig:feta} we plot the evolution of the disc inner radius assuming that the type-C QPO is the manifestation of the LTP frequency around a BH with mass 12.4$M_{\odot}$ and spin, $a
^{*}$, between 0.68 and 0.99. At the same time, the feedback fraction, $f_{\eta}$, stays at a maximum value of $\sim$ 1, which means that the disc is mainly heated up by up-scattered photons, for QPO frequencies above 2 Hz and then drops abruptly to zero at frequencies below 2 Hz, where the measured lags also change sign. For the rest of the parameters, $kT_e$, $\tau$ and $kT_s$, we did not find significant correlations with QPO frequency, except in the case of the corona temperature which exhibits, on average, lower values at frequencies below 2 Hz than above 2 Hz. More specifically, the corona temperature, $kT_e$, remains above 70 keV, on average, at frequencies above 2 Hz, while at 1.7 Hz it becomes $\sim$ 60 keV and decreases down to about 11 keV at 1.2 Hz. The amplitude of the external heating rate, $|\delta \dot{H}_{ext}|$, which is derived through comparison with the measured rms-frequency spectrum, decreases from around $10\%$ at 5.9 Hz, to $20\%$ at 2 Hz. Below 2 Hz, $|\delta \dot{H}_{ext}|$ becomes extremely high (>100$\%$), a fact that will be discussed later in section \ref{s:caveats}.

\begin{table*}
\caption{Physical parameters of the model at different frequencies of the type-C QPO in GRS $1915+105$. The frequencies in the first column represent the center of the selected bins in which the distribution of each parameter was analysed. The amplitude of the external heating rate (last column) is deduced by comparison of the simulated rms-frequency spectra to the measured one (see Figure \ref{fig:rms_sim_DHext}). The uncertainties of the model parameters were calculated as described in Figure \ref{fig:rms_sim_DHext} and are omitted when only upper limits for the parameters could be constrained.} 
\label{tab:frequency-only-approach}
\begin{tabular}{cccccccccc}
\hline
   $\nu_{qpo}$ [Hz] & $L$ [$R_g$] & $f_{\eta}$& $kT_e$ [keV] & $kT_s$ [keV] & $\tau$  & $\Gamma$ & $|\delta \dot{H}_{ext}| [\%]$\\
\hline

0.9 & 157$^{+56}_{-113}$ & 0 & 15$^{+24}_{-0}$ & 0.4$^{+1.2}_{-0}$ & 2.3$^{+2.0}_{-0}$ & 1.9$\pm0.3$ & 256.2 \\
\hline
1.2 & 185$^{+38}_{-75}$ & 0 & 11$^{+49}_{-0.}$ & 0.3$^{+1.1}_{-0}$ & 1.7$^{+2}_{-0}$ & 1.5$^{+0.5}_{-0}$ & 310.6 \\ 
\hline
1.5 & 231$^{+0}_{-64}$ & 0 & 16$^{+44}_{-9}$ & 0.2$^{+1.6}_{-0}$ & 1.2$^{+2.2}_{-0}$ & 1.5$^{+0.7}_{-0}$ & 432.2 \\ 
\hline
1.7 & 7.6$^{+84}_{-0}$ & 0.1$^{+0.5}_{}$ & 57$^{+29}_{-34}$ & 0.2$^{+2.0}_{-0}$ & 1.0$^{+1.6}_{-0}$ & 1.5$^{+0.7}_{-0}$ & 36.5 \\ 
\hline
2.0 & 7.2$^{+34}_{-0}$ & 1$^{}_{-0.2}$ & 93$^{+5}_{-29}$ & 0.3$^{+1.2}_{-0.1}$ & 1.0$^{+0.7}_{-0.0}$ & 1.6$^{+0.5}_{-0.2}$ & 21.1 \\ 
\hline
2.3 & 7$^{+43}_{-0}$ & 1$^{}_{-0.2}$ & 76$^{+22}_{-31}$ & 0.3$^{+1.2}_{-0.1}$ & 1.0$^{+1.4}_{-0.0}$ & 1.7$^{+0.4}_{-0.2}$ & 19.8 \\ 
\hline
2.6 & 30$^{+34}_{-23}$ & 1$^{}_{-0.2}$ & 79$^{+19}_{-28}$ & 0.4$^{+1.1}_{-0.3}$ & 1.0$^{+0.8}_{-0.0}$ & 1.9$^{+0.3}_{-0.4}$ & 19.1 \\ 
\hline
2.8 & 44$\pm36$ & 1$^{}_{-0.2}$ & 93$^{+5}_{-25}$ & 0.6$^{+1.0}_{-0.4}$ & 1.5$\pm0.6$ & 1.8$^{+0.4}_{-0.3}$ & 18.8 \\ 
\hline
3.1 & 58$\pm38$ & 1$^{}_{-0.2}$ & 100$^{}_{-26}$ & 0.5$^{+1.2}_{-0.3}$ & 1.5$\pm0.6$ & 1.7$^{+0.5}_{-0.2}$ & 18.5 \\ 
\hline
3.4 & 75$\pm 40$ & 1$^{}_{-0.2}$ & 76$^{+22}_{-27}$ & 0.7$^{+1.1}_{-0.5}$ & 1.1$^{+0.9}_{-0}$ & 1.7$^{+0.4}_{-0.2}$ & 18.3 \\ 
\hline
3.7 & 79$^{+39}_{-52}$ & 1$^{}_{-0.2}$ & 100$^{}_{-27}$ & 0.9$^{+0.9}_{-0.6}$ & 1.0$^{+0.8}_{-0}$ & 2$^{+0.3}_{-0.4}$ & 17.8 \\ 
\hline
3.9 & 96$\pm42$ & 1$^{}_{-0.2}$ & 74$^{+23}_{-29}$ & 0.6$^{+1.3}_{-0.3}$ & 1$^{+1.4}_{-0}$ & 1.7$^{+0.4}_{-0.3}$ & 17.1 \\ 
\hline
4.2 & 107$\pm42$ & 1$^{}_{-0.2}$ & 100$^{}_{-29}$ & 1$^{+1}_{-0.7}$ & 1$^{+1.4}_{-0}$ & 1.7$^{+0.5}_{-0.2}$ & 16.2 \\ 
\hline
4.5 & 89$^{+44}_{-73}$ & 1$^{}_{-0.2}$ & 91$^{+6}_{-24}$ & 0.9$^{+1.2}_{-0.5}$ & 0.9$^{+1.3}_{-0.0}$ & 1.9$\pm 0.3$ & 15.0 \\ 
\hline
4.8 & 101$^{+44}_{-59}$ & 1$^{}_{-0.2}$ & 100$^{}_{-24}$ & 1$^{+1.1}_{-0.7}$ & 1.1$^{+1}_{-0}$ & 1.8$^{+0.4}_{-0.3}$ & 13.6 \\ 
\hline
5.0 & 116$\pm44$ & 1$^{}_{-0.2}$ & 91$^{+6}_{-24}$ & 1.4$\pm 0.9$ & 1.0$^{+1.4}_{-0}$ & 1.6$^{+0.6}_{-0.1}$ & 12.2 \\ 
\hline
5.3 & 115$^{+45}_{-60}$ & 1$^{}_{-0.2}$ & 79$^{+18}_{-30}$ & 0.8$^{+1.2}_{-0.5}$ & 1$^{+0.8}_{-0}$ & 1.8$^{+0.4}_{-0.3}$ & 11 \\ 
\hline
5.6 & 117$^{+44}_{-59}$ & 1$^{}_{-0.2}$ & 73$^{+24}_{-30}$ & 1.3$\pm 0.9$ & 1$^{+0.8}_{-0}$ & 2$\pm 0.3$ & 10.2 \\ 
\hline
5.9 & 116$^{+45}_{-74}$ & 1$^{}_{-0.2}$ & 79$^{+18}_{-30}$ & 2.9$^{+0}_{-0.9}$ & 1.1$^{+0.9}_{-0}$ & 1.8$^{+0.4}_{-0.3}$ & 10.1 \\
\hline

\end{tabular}
\end{table*}

In the energy-dependent-approach, we fitted the rms- and lag-energy spectra of the QPO at the individual frequencies selected by \cite{Zhang2020MNRAS}. This allows us to better constrain the physical parameters of the model. This approach is equivalent to taking a snapshot of what the parameters would be at the time when the QPO frequency had a particular value. Based on the results of the energy-dependent-approach, we find that the corona size stays above 100 $R_g$ when the QPO frequency is above 2.5 Hz, while at around 2 Hz it becomes 8.7 $R_g$ with a lower $1-\sigma$ uncertainty of 6 $R_g$, indicating that around that frequency the corona size, within the uncertainties, can be as low as the lowest value in our simulations, which was 2 $R_g$. However, we note here that, based on the MCMC, the best-fitting corona size between 2 Hz and 3.5 Hz does not seem to agree with the smooth decrease in size that we derive from the frequency-only approach. This happens because the observations between 2 Hz and 3.5 Hz, for which we had energy-dependent measurements (white circles in Figure \ref{fig:plag_clustering}), happened to be selected such that the phase lags were very low, compared to other observations in the same frequency range, which naturally leads to larger fitted corona sizes, $L$, since that parameter is mostly affected by the magnitude of the phase lag. The feedback fraction, $f_{\eta}$, decreases with QPO frequency smoothly from almost 100$\%$, at 5 Hz to almost zero at 0.5 Hz. The corona temperature, although badly constrained, is high (>40 keV) at QPO frequencies above 2 Hz, and it becomes significantly lower (<13 keV) at QPO frequencies below 2 Hz. The optical depth of the corona shows a systematic increase from 0.7 at 5 Hz to 6.2 at 0.5 Hz. The seed-photon source temperature also shows a decrease, on average, from values above 1 keV at QPO frequencies above 2 Hz, to values below 0.3 keV at QPO frequencies below 2 Hz. Finally, the amplitude of the external heating rate increases from $17\%$ at 5 Hz to $\sim 40\%$ at 2 Hz, and then it becomes higher than $100\%$ at QPO frequencies below 2 Hz.

The last column of Table \ref{tab:energy-dependent-approach} summarizes the $\chi^2$ of the fits to the rms- and lag-energy spectra at each QPO frequency. For the calculation of $\chi^2$ we combined both the rms amplitude and phase lags, using the same approach as in \cite{Karpouzas2020}. The high $\chi^2$ values are, in most cases, because of the bad fit of the model to the rms-energy spectrum at energies higher than 20 keV. Although, as we mentioned before, the measured rms in energy bands above 25 keV was not used in the MCMC fitting, the rms amplitude, predicted by the model, fails to flatten out at high energies. The non-flattening of the model rms amplitude at high energies was already noticed in the case of NS LMXBs (\citealt{Karpouzas2020}), and attributed to a lack of a non-oscillating flux component in the model, such as reflection, that would contribute to a reduction of the rms amplitude at higher energies. This explanation is also valid in the case of GRS 1915$+$105, where reflection signatures appear in the energy spectrum when the type-C QPOs are observed (\citealt{Misra2020ApJ}).

\begin{table*}
\caption{Physical parameters of the model as a function of the type-C QPO frequency in GRS 1915$+$105. The best-fitting model parameters are derived via MCMC fitting to the energy-dependent rms amplitude and phase lags at each frequency,  while the corresponding uncertainties where taken as the 1$-\sigma$ confidence interval around the mode of the MCMC posterior distributions for each model parameter. The values of $\Gamma$ in the seventh column are derived using the best-fitting temperature, $kT_e$, and optical depth, $\tau$, while the uncertainties of $\Gamma$ are computed through propagation, using the uncertainties of the same parameters.}
\label{tab:energy-dependent-approach}
\begin{tabular}{ccccccccccc}
\hline
   $\nu_{qpo}$ [Hz] & $L$ [$R_g$] & $f_{\eta}$& $kT_e$ [keV] & $kT_s$ [keV] & $\tau$  & $\Gamma$ & $|\delta \dot{H}_{ext}| [\%]$ & $\chi^2/d.o.f$\\
\hline

0.466 & 1348$^{+45}_{-56}$ & 0.01$\pm 0.004$ & 7.1$^{+0.2}_{-0.3}$ & 0.2$\pm 0.05$ & 6.2$^{+0.3}_{-0.3}$ & 2.0$\pm 0.1$ & 995$^{+2}_{-4}$ & 880.9/5 \\
\hline
0.865 & 479$^{+55}_{-81}$ & 0.3$\pm 0.1$ & 7.7$^{+1.0}_{-0.8}$ & 0.2$\pm 0.04$ & 4.6$^{+0.4}_{-0.6}$ & 2.3$\pm 0.2$ & 926$^{+62}_{-600}$ & 482.4/5 \\
\hline
1.334 & 208$^{+38}_{-42}$ & 0.4$\pm 0.1$ & 12$\pm 2.4$ & 0.1$\pm 0.05$ & 3.6$\pm 0.7$ & 2.3$\pm 0.4$ & 826$^{+127}_{-218}$ & 54.6/5 \\
\hline
1.651 & 161$^{+25}_{-29}$ & 0.4$\pm 0.1$ & 13$^{+3}_{-2}$ & 0.1$\pm 0.03$ & 4.0$\pm 0.7$ & 2.1$\pm 0.3$ & 719$\pm 187$ & 153.9/5 \\
\hline
2.077 & 9$^{+8}_{-6}$ & 0.7$\pm 0.1$ & 111$^{+51}_{-43}$ & 1.3$\pm 0.3$ & 2.3$^{+1.3}_{-1.1}$ & 1.3$^{+0.2}_{-0.3}$ & 39$^{+7}_{-10}$ & 24.3/5 \\
\hline
2.538 & 138$^{+31}_{-28}$ & 0.9$\pm 0.1$ & 86$^{+52}_{-40}$ & 1.3$\pm 0.2$ & 1.0$^{+0.6}_{-0.4}$ & 2.1$^{+0.6}_{-0.8}$ & 21$\pm3$ & 7.0/5 \\
\hline
3.137 & 136$^{+31}_{-34}$ & 0.9$\pm 0.1$ & 76$^{+42}_{-45}$ & 1.2$^{+0.4}_{-0.3}$ & 1.1$\pm 0.5$ & 2.1$\pm 0.8$ & 21$\pm 5$ & 17.6/5 \\ 
\hline
4.097 & 96$^{+10}_{-14}$ & 0.8$\pm 0.04$ & 40.3$^{+18.1}_{-14.2}$ & 1.0$\pm 0.1$ & 1.2$^{+0.4}_{-0.3}$ & 2.7$^{+0.6}_{-0.8}$ & 17.3$^{+1.8}_{-2.1}$ & 11.2/5 \\
\hline
4.979 & 130$^{+15}_{-22}$ & 0.99$\pm 0.01$ & 69$^{+27}_{-16}$ & 2.9$^{+0.1}_{-0.2}$ & 0.7$\pm 0.2$ & 2.7$\pm 0.6$ & 17$^{+1}_{-2}$ & 42.4/5 \\
\hline

\end{tabular}
\end{table*}


\begin{figure}
\begin{center}

	\includegraphics[width=10cm,height=14cm]{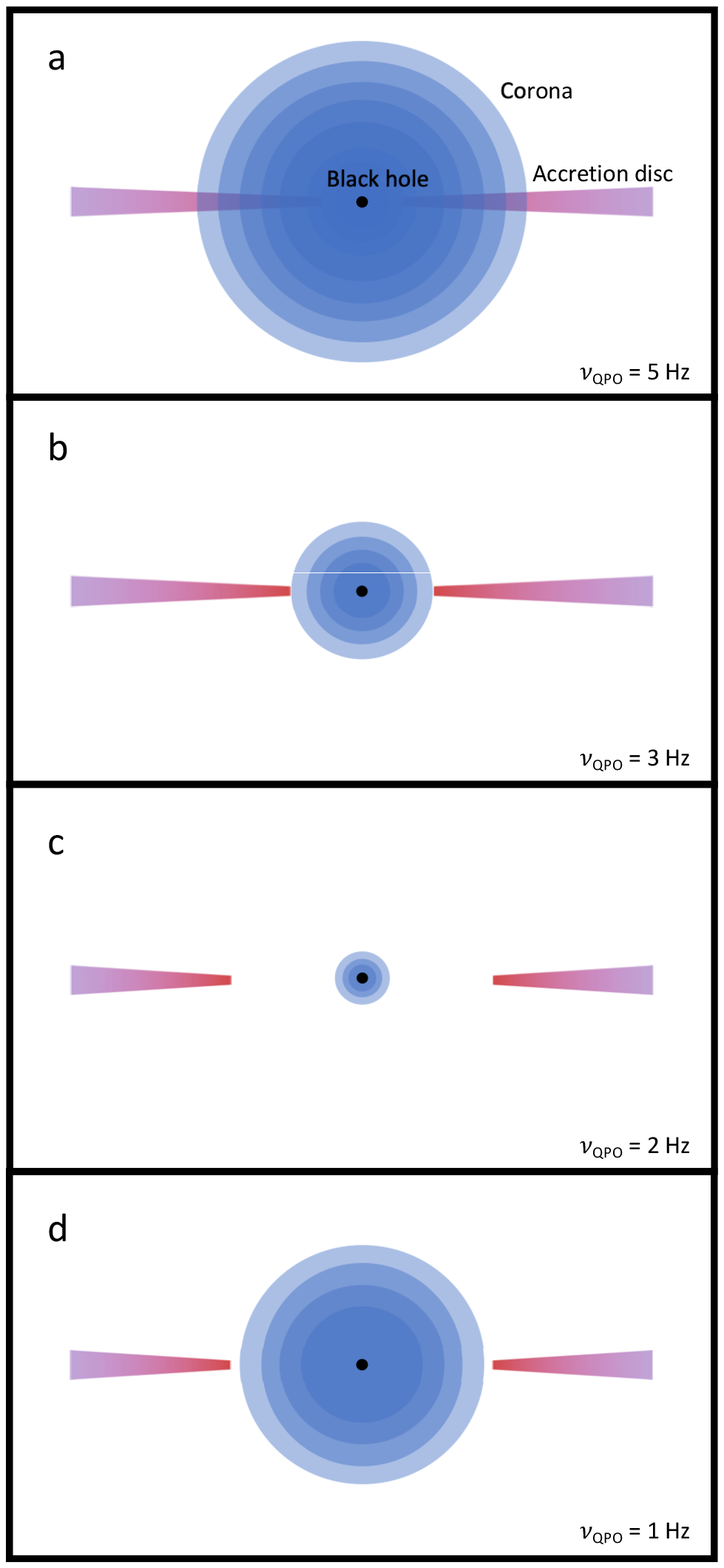}
    \caption{ Schematic representation of the evolution of the corona size in GRS 1915$+$105 at four typical QPO frequencies. In panel a, at high QPO frequency, a spherically symmetric corona (blue layered circle) covers the inner accretion disc (thick horizontal line) and feedback is very efficient. In panel b the QPO frequency is lower and, assuming it reflects the Lense-Thirring frequency at the inner edge of the disc, the inner-disc radius increases; the size of the corona has decreased enough so that it is only marginally covering the inner disc, while feedback efficiency has decreased. In panel c, where the QPO frequency is at around 2 Hz, the inner-disc radius is higher, the corona has reached its minimum size and is now inside the disc. Finally, in panel d the corona size increases again as the QPO frequency decreases, and the inner edge of the disc increases further.}
    \label{fig:schematic}
\end{center}
\end{figure}

\section{Discussion}

We propose a novel explanation for the change of the time lags of the type-C QPO in GRS 1915$+$105 from soft, at high QPO frequencies, to hard, at low QPO frequencies. We applied a Comptonisation model that incorporates feedback of the hard corona photons onto the soft photon source to a large data set of timing measurements of this source with the RXTE satellite. We used the rms amplitude and phase lag measurements of \cite{Zhang2020MNRAS} combined with the model of \cite{Karpouzas2020} and conclude that a corona with time-varying size, accompanied by a switch in feedback efficiency, is able to explain the data. In particular, we find that, as the QPO frequency decreases from 5 Hz to 2 Hz, the size of the corona decreases from around 120 $R_g$ to a few $R_g$. During this contraction, the accretion disc is covered by the corona, the feedback of Compton up-scattered photons onto the disc is very efficient ($\sim$ 100$\%$), and most of the disc flux is due to feedback photons. As the QPO frequency decreases, so does the feedback efficiency because the corona covers less and less area of the disc, as its size decreases. At the same time, assuming that the truncation radius of the disc is anti-correlated to the QPO frequency, the inner radius of the disc increases and at some critical frequency ($\sim 1.8$ Hz) the disc is no longer covered by the corona. From that point on, the efficiency of feedback decreases as the disc inner radius moves further out and the corona size continues to decrease. In Figure \ref{fig:schematic}, we present a schematic representation for the evolution of the size of the corona in four phases as the type-C QPO frequency decreases from 5 Hz down to 1 Hz. Our fits are systematically worse when the QPO frequency is below 1.8 Hz than when it is above that value. Our results indicate that when the QPO frequency is below 1.8 Hz the size of the corona increases to values much larger than 130 $R_g$, the feedback efficiency decreases to values <50$\%$ and down to zero, and also the power of the source of external heating for the corona oscillates with an amplitude larger than $100 \%$. The aforementioned paradigm of a decreasing corona and feedback efficiency fits well with all available timing measurements of the type-C QPO in the frequency range of 1.8$-$5 Hz, and explains both the decrease in phase lags between two energy bands, and the smooth decrease in the slope of the lag-energy spectrum. The change in the behaviour of the lags at 1.8 Hz and below indicates either a possible switch in the physical mechanism that produces the lags or a change in the geometry of the Comptonising region that the model does not take into account.

\subsection{Comparison to other models}

 \cite{Nobili2000ApJ} proposed a model that was able to explain the lag-frequency spectrum of GRS 1915$+$105. The difference between their approach and ours is in the mechanism that produces the soft lags. In the model of \cite{Nobili2000ApJ}, the soft lags are caused by Compton down-scattering of soft disc photons that have previously suffered saturated Comptonisation in an inner optically thick component of the corona. In the model of \cite{Nobili2000ApJ} the Compton down-scattering happens in an outer cooler component of the corona. 

\par In the model of \cite{Karpouzas2020}, the soft lags are interpreted as a delayed response of the seed photon source to feedback photons that impinge back on the disc after being Compton up-scattered in the corona. The differences in the physical assumptions between the two models are reflected on the correlation between the inferred corona size and disc inner radius. In the model of \cite{Nobili2000ApJ}, the disc inner radius is positively correlated to the size of the corona, whereas in our model the two quantities are anticorrelated down to the limit of $\sim$2 Hz and then positively correlated for lower QPO frequencies. Our model, however, explains not only the lag-frequency spectrum, but also both the rms- and lag-energy spectra at different QPO frequencies. 

\par \cite{Dutta2018MNRAS} based on the TCAF model, provide an explanation opposite to ours when it comes to the evolution of the Comptonising cloud at different type-C QPO frequencies. More specifically, although the size of our corona is of the same order of magnitude as the size of the CENBOL in \cite{Dutta2018MNRAS}, in their interpretation there is no systematic or rapid decrease in the CENBOL size at around 1.8 Hz. Since a transition of a type-C QPO frequency around 1.8 Hz typically happens in matter of a few days (\citealt{Dutta2018MNRAS}), the predicted decrease in the size of the corona should happen on the same time scale. Regardless of whether the decrease in corona size is systematic or very rapid, our fits to both the lag-energy and lag-frequency spectra agree that the size should be significantly lower around 1.8 Hz. If our explanation is correct, a reduction in the size of the corona, while the temperature remains the same or decreases, would lead to a reduction in the hard X-ray intensity at low QPO frequencies. \cite{Stiele2018ApJ} and \cite{Bhargava2019MNRAS} provide evidence of positive correlation between QPO frequency and X-ray flux. Furthermore, if matter from the corona is not advected onto the black hole then, given appropriate conditions in the magnetic field, matter could be directed towards the jet, in which case enhanced radio emission should be present at low QPO frequencies. In support to our expectations, \cite{Yan2013MNRAS}, found that increased radio activity is present when the frequency of the type-C QPO is low.

\par Although at first glance our explanation appears to be in conflict with the model of \cite{Ingram2009MNRAS}, this is not the case. \cite{Ingram2009MNRAS} assumed that the inner flow around an accreting black hole consists of a geometrically thick torus located inside the truncation radius of a non-precessing geometrically thin disc. As the torus precesses, it illuminates different parts of the disc causing the modulation of the X-ray flux that produces the LF QPO. \cite{Ingram2016} noted, however, that the same pattern would result if the torus was fixed and it was the disc the one that precessed at the LT frequency (\citealt{Schnittman2006ApJ}; \citealt{Tsang2013MNRAS}). As \cite{Ingram2016} explained, their choice of one geometry over the other was based on the fact that the rms spectrum of the QPO is hard and hence the emission at the QPO frequency could not come from the disc.

\par  As we showed here, this argument is not applicable, and a scenario with a precessing disc and a fixed corona is possible. Indeed, in our model the rms spectrum of the QPO is a consequence of inverse Compton scattering of soft disc photons in the corona (the torus in the scenario of \citealt{Ingram2009MNRAS}), such that the high rms amplitude values of the QPO at high energies simply reflect the variability of the soft disc emission at the LT frequency that is scattered in the corona. This, plus the feedback from the corona to the disc, naturally explain both the variability of the iron line discussed in \cite{Ingram2016} and the rms spectrum of the QPO. 

Our interpretation of a corona with a different size at different QPO frequencies, agrees with recent evidence presented by \cite{Kara2019Natur} of a contracting corona in the BH transient MAXI J1820$+$070. \cite{Kara2019Natur} studied the broad-band noise component through a reverberation model that assumes a point-like corona above the accretion disc, and presented evidence that the height of the corona changes with time. Here, we used the LF QPOs and applied a Comptonisation model that assumes an extended corona, and provided evidence that the size of this corona changes with QPO frequency. Since the QPO frequency changes with time, the corona size also changes with time. The results of \cite{Kara2019Natur} and ours could be due to a similar mechanism that acts throughout the different spectral states and depending on the exact state, and its spectral-timing properties can be detected by using different analysis and modelling techniques.    

\subsection{On the possible connection between the corona and the jet}

 Persistent and transient BH LMXBs exhibit significant radio emission (\citealt{Fender2001MNRAS}). The radio activity is related to the existence of a jet (\citealt{Brand1986ApJ}; \citealt{Levinson1996ApJ}; \citealt{Georganopoulos2002A&A}; \citealt{Reig2003A&A}), that can produce hard X-rays through Compton up-scattering. Jet models, such as the ones presented in \cite{Reig2003A&A} and \cite{Kylafis2018A&A}, successfully explain the hard lags as a function of broad-band frequency in the context of Comptonisation of disc seed-photons in the jet. However, to the best of our knowledge, the soft lags of the type-C QPOs, and particularly the transition from hard to soft lags, has yet to be explained by jet models. 
 
 \par The radio emission in GRS 1915$+$105 is occasionally anti-correlated to the hard X-ray flux (\citealt{Trudolyubov2001ApJ}). The fact that the radio emission of GRS 1915$+$105 is stronger at lower type-C QPO frequencies (\citealt{Yan2013MNRAS}), combined with the fact that the lags are hard at low QPO frequencies and, as we showed here, Comptonisation in a uniform corona provides systematically worse fits to the data, naturally leads to the following scenario: At high QPO frequencies an extended corona with efficient feedback onto the disc can explain both the power-law component in the energy spectrum and the lags and rms-amplitude of the type-C QPO. As the size of the corona decreases, feedback onto the disc and Comptonisation in the corona are no longer the dominant components that drive the timing-spectral properties, and so the jet takes over as the medium that produces the power-law and the hard lags. This scenario, if validated, can explain all the spectral and timing properties at once, as well as the correlation between X-ray and radio luminosity.

\subsection{Model Caveats}
\label{s:caveats}

We modelled the rms-frequency, lag-frequency, rms-energy and lag-energy spectra of the type-C QPO in GRS 1915$+$105, using the Comptonisation model of \cite{Karpouzas2020}. As with every model, there are certain caveats that we address here. Firstly, the model tends to produce less simulated phase-lag and rms-amplitude values at QPO frequencies lower than 1.8 Hz (Figures \ref{fig:plag_clustering} and \ref{fig:rms_clustering}). This happens because in the model the energy-dependent rms-amplitude and lags do not change as smoothly for low values of the feedback fraction, $f_{\eta}$, as for high values ($f_{\eta}$>0.3). This behaviour is inherent to the model and depends on the assumptions upon which it is built. To produce hard lags, the model requires less feedback, since in the model increasing the feedback means increasing the time-delay of the soft component with respect to the hard one, and so regimes of hard lags will tend to be poorly sampled. A detailed study of how the model depends upon each physical parameter will be presented in an upcoming publication (Garc\'ia et al. in prep.).

\par Secondly, the model of \cite{Karpouzas2020} uses a simple blackbody as the seed-photon source for Comptonisation. Despite this simplification, it is remarkable how good the model works when applied to the data. While a multi-colour disc-blackbody would be a more realistic assumption, this would not only add two additional free parameters (source inclination and disc inner radius), but it would complicate the definition of both the optical depth, which should then have an extra spatial dependence, and of the feedback, since the feedback efficiency should then depend on the inner disc radius. Addressing the aforementioned problems will be part of a future work, especially since source inclination is suspected to play a role in the switch of the time lags behaviour as a function of QPO frequency (\citealt{vde2017MNRAS}).

\par Finally, the model yields a seemingly worse fit to the data of the rms-energy spectra above 25 keV, and also in general at QPO frequencies below 1.8 Hz. The first issue was also addressed in \cite{Karpouzas2020} and is likely due to the lack of a reflection component in our model. The second issue, which becomes even more apparent from the very large values of the required external heating rate amplitude ($|\delta \dot{H}_{ext}|$>100$\%$), can be due to either a break-down of the assumption of the model at QPO frequencies below 1.8 Hz, or to an actual switch in the physical mechanism, as mentioned in the previous sub-section. To answer as many of these questions as possible and retain applicability to large data-sets, a more detailed modeling of the disc-corona-jet geometry and interaction in a semi-analytical framework is needed.

\section{Conclusions}

We applied our newly developed Comptonisation model, initially designed to explain the lower kHz QPO in NS LMXBs, to the type-C QPO of the BH LMXB GRS 1915$+$105. We used a large data-set of timing measurements of this source obtained using archival data from the RXTE satellite. We showed that a spherically symmetric and uniform corona of a finite size is able to explain these timing properties, under the condition that the size of the corona decreases as a function of QPO frequency, down to the critical frequency of 1.8 Hz, where the lags turn from soft to hard. Furthermore, the switch of the lags from soft to hard is moderated by the feedback efficiency of up-scattered photons onto the accretion disc that lies inside the corona. For the first time, we were able to fit the energy dependent rms amplitude and time lags of type-C QPOs, which supports the idea that Comptonisation in the corona should be an essential component of every spectral and timing model.

\section*{Acknowledgements}
 D.A acknowledges support from the Royal society. This work is part of the research programme Athena with project
number 184.034.002, which is (partly) financed by the Dutch Research
Council (NWO). LZ acknowledges support from the Royal Society Newton Funds. TMB acknowledges financial contribution from the agreement ASI-INAF n. 2017-14-H.0. Y.Z. acknowledges the support from China Scholarship Council (CSC 201906100030). This research has made use of data obtained from the High Energy Astrophysics Science Archive Research Center, provided by NASA's Goddard Space Flight Center.






\bibliographystyle{mnras}
\bibliography{references} 

\bsp	
\label{lastpage}
\end{document}